\documentclass[prb,aps,showpacs,twocolumn,amsmath,amssymb,floatfix,superscriptaddress]{revtex4-2}

\usepackage[unicode=true, colorlinks=true, citecolor={blue!80!black}, urlcolor={blue!50!black}, linkcolor = {blue!80!black}]{hyperref}
\usepackage{graphicx}
\usepackage{bm}
\usepackage{dsfont}
\usepackage{xcolor}
\usepackage[utf8]{inputenc}
\usepackage{amssymb}
\usepackage[normalem]{ulem}
\usepackage{physics}

\newcommand{\beq}{\begin{equation}}
\newcommand{\eeq}{\end{equation}}
\newcommand{\beqa}{\begin{eqnarray}}
\newcommand{\eeqa}{\end{eqnarray}}

\definecolor{quote-color}{rgb}{1.00, 0.30, 0.30}

\begin{document}

\title{Josephson effect and critical currents in trivial and topological full-shell hybrid nanowires}
%\title{JMajorana modes revealed by critical currents in Josephson junctions based on full-shell hybrid nanowires}
\author{Carlos Payá}
\affiliation{Instituto de Ciencia de Materiales de Madrid (ICMM), CSIC, Madrid, Spain}
\author{Ramón Aguado}
\affiliation{Instituto de Ciencia de Materiales de Madrid (ICMM), CSIC, Madrid, Spain}
\author{Pablo San-Jose}
\affiliation{Instituto de Ciencia de Materiales de Madrid (ICMM), CSIC, Madrid, Spain}
\author{Elsa Prada}
\email{elsa.prada@csic.es}  
\affiliation{Instituto de Ciencia de Materiales de Madrid (ICMM), CSIC, Madrid, Spain}

\date{\today}

\begin{abstract}
We perform microscopic numerical simulations of the Josephson effect through short junctions between two full-shell hybrid nanowires, comprised of a semiconductor core fully wrapped by a thin superconductor shell, both in the trivial and topological regimes.
We explore the behavior of the current-phase relation and the critical current $I_c$ as a function of a threading flux for different models of the semiconductor core and different transparencies of the weak link.
We find that $I_c$ is modulated with flux due to the Little-Parks (LP) effect and displays a characteristic \textit{skewness} towards large fluxes within non-zero LP lobes, which is inherited from the skewness of a peculiar kind of subgap states known as Caroli--de Gennes--Matricon (CdGM) analogs.
The appearance of Majorana zero modes at the junction in the topological phase is revealed in $I_c$ as \textit{fin}-shaped peaks that stand out from the background at low junction transparencies. The competition between CdGMs of opposite electron- and hole-like character produces steps and dips in $I_c$. A rich phenomenology results, which includes 0-, $\pi$- and $\phi$-junction behaviors depending on the charge distribution across the wire core and the junction transparency.
\end{abstract}

\maketitle

\section{Introduction}

%Although the phenomenology of single full-shell wires has been extensively analyzed, a characterization of Josephson junctions in the topological phase is still lacking. In this work, we consider the short-junction stationary Josephson effect between two such full-shell hybrid nanowires.
%We explore the Josephson effect through a superconductor-normal-superconductor (SNS) junction based on full-shell hybrid wires in the short-junction limit.

The Josephson effect \cite{Josephson:PL62,Josephson:RMP74,Barone:82} --a macroscopic quantum phenomenon that occurs when two superconductors are brought into proximity through a tunnel barrier or a weak link-- is a central concept in the physics of superconductors. It has numerous applications, e.g. in metrology, quantum computing, and quantum sensing \cite{Barone:82,Tinkham:96,Gross:16,Aguado:24}. For example, it is at the heart of superconducting qubits \cite{Kjaergaard:ARCMP20}, one of the most promising platforms for quantum computing. Josephson junctions have been extensively studied in recent years in hybrid superconductor-semiconductors platforms \cite{Larsen:PRL15, deLange:PRL15, Hays:PRL18, Aguado:APL20, Larsen:PRL20, Hays:S21, Pita-Vidal:NP23,Matute-Canadas:PQ24} due to the versatility and tunability these systems offer compared to metallic junctions, ranging from the early demonstration of gate-tunable supercurrents \cite{Doh:S05}, to heterojunctions \cite{Schiela:PQ24} or the recent demonstrations of diode effect in interferometers \cite{Ciaccia:PRR23,Reinhardt:NC24}, just to name a few. Furthermore, in this kind of platforms it is in principle possible to engineer a topological superconducting phase \cite{Beenakker:ARCMP13,Aguado:RNC17,Lutchyn:NRM18,Prada:NRP20,Marra:JoAP22}, opening the door to the realization of topologically protected qubits based on Majorana zero modes (MZMs) \cite{Nayak:RMP08,Aasen:PRX16,Karzig:PRB17,Aguado:PT20,Beenakker:SPLN20}.

Within the subfield of hybrid nanowires, sometimes dubbed Majorana nanowires \cite{Lutchyn:PRL10, Oreg:PRL10}, an alternative geometry with several advantages was introduced five years ago \cite{Vaitiekenas:S20,Penaranda:PRR20,Kopasov:PRB20,Kopasov:PSS20}. These are called full-shell hybrid nanowires, a semiconductor nanowire with strong spin-orbit coupling (SOC) covered all around by a thin superconductor shell, in contrast to conventional Majorana nanowires, where the shell only covers some of the wire facets \cite{Krogstrup:NM15}. It has been argued that this geometry can host a topological superconducting phase driven by the orbital effect of an applied axial magnetic field \cite{Vaitiekenas:S20} (instead of by the Zeeman effect like in their partial-shell counterparts) whenever an odd number of superconductor phase windings, or fluxoids, are induced by the magnetic flux threading the hybrid section. In practice, this reduces considerably the magnitude of the magnetic field required for the topological transition. Moreover, a complete encapsulation of the semiconductor core in the metallic superconducting shell also reduces the effect of certain forms of disorder (such as the one coming from the electrostatic environment \cite{Vuik:NJP16,Escribano:BJN18}).

Most studies have focused on single full-shell hybrid nanowires, examined e.g. through tunneling or Coulomb spectroscopy experiments \cite{Valentini:S21, Valentini:N22, Valentini:PRR25,Deng:PRL25}. Theoretically, the spectrum of these wires is characterized mainly by two features. One is the Little-Parks (LP) effect \cite{Little:PRL62, Parks:PR64}  of the tubular shell, by which the superconducting order parameter oscillates with applied flux in a series of \emph{lobes} characterized by an integer number $n$ of fluxoids \cite{Vaitiekenas:PRB20,Sabonis:PRL20,Vekris:SR21}. The second is the appearance of a special type of Andreev subgap states in the semiconductor core, called Caroli-de Gennes-Matricon (CdGM) analogs  \cite{San-Jose:PRB23,Paya:PRB24} because of their similarity to CdGM states in type II vortices \cite{Caroli:PL64, BrunHansen:PLA68, Bardeen:PR69}. These states have been recently demonstrated experimentally in Al/InAs based full-shell nanowires \cite{Deng:PRL25}. Their phenomenology has been predicted to strongly depend on the electron charge distribution inside the core, which is essentially inaccessible due to the superconductor encapsulation; thus, several models are typically considered from the theoretical perspective \cite{Paya:PRB24, Vezzosi:SP25}.

%While the Josephson effect has been extensively studied in standard partial-shell nanowires \cite{San-Jose:PRL12,San-Jose:PRL14,Nesterov:PRB16,Cayao:PRB17}, studies focusing on full shell NWs are very rare and limited to simplified models [Giorgos y quizás algún otro más?]. 

%Josephson junctions and related structures \editE{have been extensively examined in partial-shell nanowires \cite{San-Jose:PRL12,San-Jose:PRL14,Nesterov:PRB16,Cayao:PRB17}, but its study} 

% \footnote{A previous work considered theoretically full-shell based Josephson junctions, but only in the idealized hollow-core approximation \cite{Giavaras:PRB24}.} 

Although the Josephson effect has been extensively examined in standard partial-shell nanowires \cite{San-Jose:PRL12,San-Jose:PRL14,Nesterov:PRB16,Cayao:PRB17,Sriram:PRB19,Zellekens:PRA20,Razmadze:PRB23,Legg:PRB23}, its study in full-shell nanowires has only recently begun to be explored, either with simplified models \cite{Giavaras:PRB24} or mostly from an experimental point of view. For example, studies have been published on Joule heating \cite{Ibabe:NC23, Ibabe:NL24}, Coulomb islands \cite{Razmadze:PRB24a}, or Aharonov-Bohm-type oscillations \cite{Zellekens:24} influencing the supercurrent of these wires. 

\begin{figure}
   \centering
   \includegraphics[width=\columnwidth]{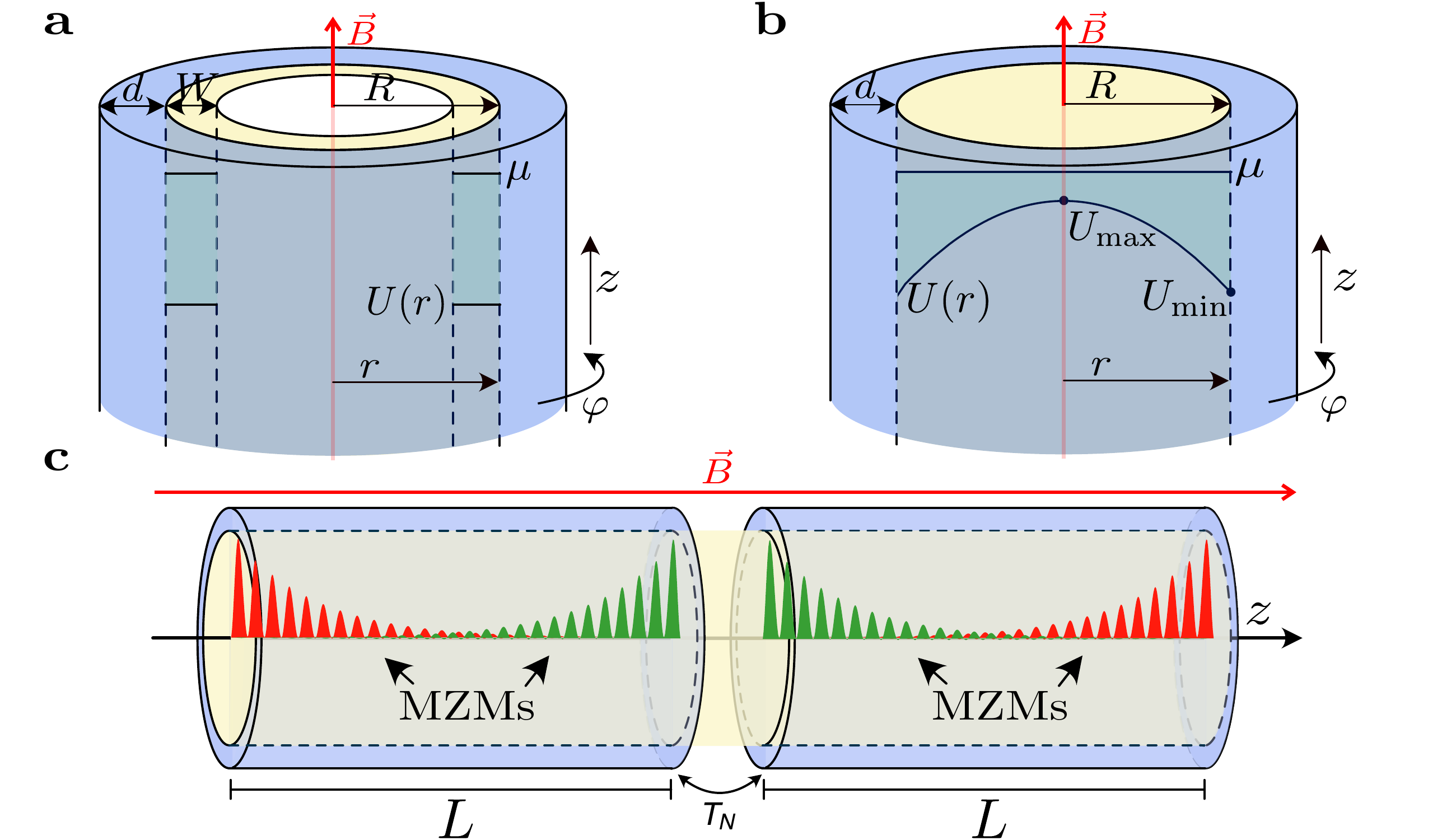}
   \caption{(a) Sketch of a full-shell hybrid nanowire with a tubular-core geometry in a cylindrical approximation. An insulating core (white) is surrounded by a semiconductor tube (yellow) of external radius $R$ and thickness $W$, and is completely encapsulated in a thin superconductor shell (blue) of thickness $d$. In an applied axial magnetic field $B$ the hybrid wire is threaded by a non-quantized magnetic flux $\Phi=\pi(R+d/2)^2B$. The chemical potential $\mu$ and the radial electrostatic potential energy $U(r)$ are schematically depicted inside. (b) Same as (a) but for a semiconductor solid-core geometry. The conduction-band bottom inside the semiconductor exhibits a dome-like radial profile with maximum value at the center, $U_{\rm{max}}$, and minimum value at the superconductor-semiconductor interface, $U_{\rm{min}}$. (c) Sketch of a weak-link Josephson junction between two full-shell nanowires of finite length $L$ in the topological phase. In red and green, Majorana zero mode (MZM) wavefunctions at the left and right ends of the nanowires along the longitudinal direction $z$. $T_{\rm{N}}$ is the normal transmission that characterizes the weak link.}
   \label{fig:sketch}
\end{figure}

In this work, we theoretically characterize for the first time the rich phenomenology of Josephson junctions based on full-shell hybrid nanowires in the zero-temperature and short-junction limits, both in the trivial and topological phases, considering different models for the semiconductor core and different transmissions of the weak link; see Fig. \ref{fig:sketch}. We find that the critical current $I_c$ through these junctions follows a periodic modulation with flux coming from the LP effect. Within each LP lobe, $I_c$ strongly depends on the behavior of the CdGM analogs, which itself depends on the distribution of the charge density across the core section. In general, this translates into a skewness of $I_c$ towards large flux values within each nonzero LP lobe, with a fine step-like structure that reflects zero-energy crossings of the CdGM analogs with flux. For wire parameters in the topological phase, the appearance of MZMs at the junction manifests itself as a conspicuous increase of the critical current at low junction transparencies (in the form of \textit{fin}-shaped peaks with flux). The Majorana fins disappear gradually as the junction transparency increases. This feature could potentially be used to detect the presence of MZMs in full-shell nanowire-based Josephson junctions through critical current measurements. Finally, the intricate behavior of CdGM analogs, especially in hybrid wires where their charge density spans the entire semiconductor core, occasionally brings the critical current to zero at specific flux values. This signals a transition from the usual $0$- to a $\pi$- or $\phi$-junction regime \cite{Barone:82, Buzdin:PRB03}.

This paper is organized as follows. In Sec. \ref{Sec:model} we introduce the different full-shell hybrid nanowire models as well as the methodology. Further details can be found in Appendix \ref{ap:hamiltonian}. In Sec. \ref{Sec:trivial} we consider the Josephson effect in short junctions with semi-infinite full-shell hybrid nanowires in the trivial regime, whereas topological junctions are studied in Sec. \ref{Sec:topological}. The case with finite-length full-shell hybrid nanowires is discussed in Sec. \ref{Sec:finite}. Finally, we conclude in Sec. \ref{Sec:conclusions}.

\section{Models and methods}
\label{Sec:model}

We study Josephson junctions based on full-shell hybrid nanowires. We employ a cylindrical description using coordinates $(r,\varphi,z)$ since it has been proven that it is a good approximation to describe more realistic hexagonal cross-section wires \cite{Paya:PRB24}. The hybrid wires are characterized by a core radius $R$, a superconductor shell thickness $d$ and length $L$, which we take as infinite in Secs. \ref{Sec:trivial} and \ref{Sec:topological}, and keep finite in Sec. \ref{Sec:finite}. In our calculations, we focus on Al/InAs hybrid nanowires. In this type of wires, there is typically a charge accumulation layer in the core close to the superconductor-semiconductor interface because of the Ohmic contact between both materials. The specific details of the electrostatic profile inside the shell are device-dependent and nontunable through external gates, due to the metallic shell encapsulation. For this reason, we consider different models for the semiconductor core. These are summarized in Appendix \ref{ap:hamiltonian} and expanded upon in Refs. \cite{San-Jose:PRB23, Paya:PRB24}.

In the tubular-core model, Fig. \ref{fig:sketch}(a), the semiconductor has a tubular shape with thickness $W$, roughly approximating the width of the accumulation layer. For simplicity, this model assumes constant values for both the chemical potential ($\mu$) and the SOC ($\alpha$) throughout the semiconductor. This model could also be used to describe a semiconductor core-shell nanowire, where a semiconductor shell surrounds an insulating core. A limiting and idealized case of this model is known as the hollow-core approximation, where $W\rightarrow 0$. Despite this approximation, this model is useful to understand the more realistic behavior of tubular- and solid-core models.

In the solid-core model, Fig. \ref{fig:sketch}(b), the semiconductor fills the entire core. This approach incorporates the conduction band bending caused by differences between the superconductor’s work function and the semiconductor’s electron affinity, modeled via an electrostatic potential $U(r)$ \cite{Mikkelsen:PRX18,Antipov:PRX18, Vaitiekenas:S20, Paya:PRB24}. The spatial gradient of $U(r)$ determines the SOC field, which we calculate using standard formulae; see Appendix \ref{ap:hamiltonian}.

For the shell, we consider a diffusive superconductor with gap $\Omega_0$ at zero magnetic field and superconducting coherence length $\xi_d$. We assume that there is a decay rate $\Gamma_S$ between the semiconductor and the superconductor. An axial magnetic field $\vec{B} = B \hat{z}$ is applied, inducing a flux $\Phi = \pi (R + d/2)^2 B$ through the hybrid wire section \footnote{Note that $\Phi$ is taken at the mean radius of the shell, $R_{\rm{LP}}=R+d/2$ \cite{Vaitiekenas:S20,Paya:PRB24}. In the thin-shell limit, $d\ll \lambda_L$ (where $\lambda_L$ is the London penetration length), the shell does not have the ability to screen the axial magnetic field, so that the applied field is the same everywhere. The associated supercurrents grow linearly with radius, like the vector potential. For small shell thickness $d\ll R$, the supercurrent can be approximated by its spatial average (i.e. its value at $R_{LP}=R+d/2$). This approximation is more accurate than simply replacing $d$ by zero in the LP equations.}. The presence of a magnetic field in a thin tubular-shaped superconductor implies a quantization of the fluxoid (not the flux) in the shell in units of the superconducting flux quantum $\Phi_0 = h/2e$ ($h$ is Planck's constant and $e$ the electron charge). The fluxoid number describes how many times the superconducting phase winds around the wire axis, and it is given by $n(\Phi) = \lfloor \Phi /\Phi_0 \rceil$ \cite{Little:PRL62, Parks:PR64}. In the presence of $B$, a modulation of the shell gap with flux appears, $\Omega(\Phi)$, giving rise to LP lobes, each characterized by different fluxoid numbers $n$; see Appendix \ref{ap:hamiltonian} and Ref. \cite{San-Jose:PRB23}. The flux modulation of the superconducting properties is a consequence of the pair-breaking effect of the magnetic field on the Cooper pairs in the superconductor. This pair-breaking effect is minimal at integer values of $\Phi/\Phi_0$, where $\Omega$ reaches a maximum, and strongest at half-integer values, where the gap is minimized. Shells with a finite (zero) $\Omega$ at this point are said to be in the non-destructive (destructive) regime.

The effective Bogoliubov-de Gennes (BdG) Hamiltonians $H_{\rm{BdG}}$ describing the three hybrid wire models can be found in Appendix \ref{ap:hamiltonian}. Cylindrical symmetry allows us to express $H_{\rm{BdG}}$ in terms of the generalized angular momentum quantum number $m_J$, that labels the different occupied transverse subbands of the hybrid wire and take values
\beq
m_J = \left\{\begin{array}{ll}
\mathbb{Z}+ \frac{1}{2} & \textrm{if $n$ is even} \\
\mathbb{Z} & \textrm{if $n$ is odd}
\end{array}\right..
\label{mJ}
\eeq
This points to qualitative differences between the spectrum in even and odd LP lobes. Other relevant parameters of the Hamiltonian are the effective mass $m^*$ and the $g$ factor.

Next, we define a weak-link Josephson junction connecting two equal full-shell hybrid nanowires; see Fig. \ref{fig:sketch}(c). We focus on short junctions with length $L_N\ll \xi_d$ \cite{Beenakker:PRL91, Beenakker:TPMS92, Beenakker:NaMS92, Golubov:RMP04, Fatemi:SP25}. Numerical calculations rely on a tight-binding description of the effective $H_{\rm{BdG}}$, with discretization lattice parameter $a_0$. We use a Green's function approach to calculate the supercurrent $J_S(\phi)$ as a function of the superconducting phase different $\phi$ across the junction. All the necessary information of the left and right superconducting segments is encoded in their respective $g^r$, the bare Green's function at the end of each of the two decoupled hybrid nanowires, expressed in terms of $H_{\rm{BdG}}$.
%We focus on short junctions with length $L_N\ll \xi_d$ and characterized by an inter-wire normal transmission $T_N$.}
%\editE{In turn, this tight-binding discretization of $H_{\rm{BdG}}$ allows us to include all the microscopic information of each isolated full-shell L/R segment, see Fig. \ref{fig:sketch}(c), within a Green's function approach. Since the segments are semi-infinite, this information is fully encapsulated in $g^r$, the bare Green's function at the end of the hybrid nanowire. Next, we define a weak-link Josephson junction connecting both equal segments. We focus on short junctions with length $L_N\ll \xi_d$.}
Observables are calculated using the retarded dressed Green's function within the first unit cell of one of the two wires, $G^r = \left[\left(g^r\right)^{-1} - \Sigma^r\right]^{-1}$, where $\Sigma^r = V^\dagger g^r V$ is the tunneling self-energy from the other wire defined in terms of the coupling matrix $V$ between them. We take $V$ as a fraction of the intra-wire hopping term between unit cells. Physically, $V$ defines an inter-wire normal transmission $T_N$.

Using this framework, the local density of states (LDOS) at either side of the junction for energy $\omega$ and phase difference $\phi$ is given by
\begin{equation} \label{eq:LDOS}
    \rho(\omega, \phi)= -\frac{1}{\pi}\text{Im}\left[\text{Tr}\left(G^r\right)\right].
\end{equation}
The equilibrium Josephson current, formulated within the Keldysh formalism \cite{Abrikosov:63, Perfetto:PRB09, Martin-Rodero:PRL94, Sun:PRB00}, is expressed as
\begin{equation} \label{eq:josK}
     J_S(\phi) = \frac{2e}{h} 2\text{Re} \int d\omega f(\omega) \text{Tr}\left[\left(\Sigma^rG^r - G^r\Sigma^r\right)\tau_z\right],
\end{equation}
where $f(\omega)$ is the Fermi-Dirac distribution, traces are taken over spin and electron-hole degrees of freedom and $\tau_z$ is the a Pauli matrix in electron-hole space. We have written it in terms of retarded Green’s functions, guaranteeing the analyticity of the integrand in the complex plane, which allows an efficient numerical evaluation along a complex $\omega$-path~\cite{Rakyta:PRB16}. Note that in all our calculations we take $T\rightarrow 0$, although Eq. \eqref{eq:josK} allows to simulate finite temperatures at no extra cost.

In this work, we calculate $J_S(\phi)$ using Eq. \eqref{eq:josK}, which is numerically efficient, but it is instructive to note that, alternatively, the Josephson current can also be written as
\begin{equation} \label{eq:josF}
    J_S(\phi) = \frac{2e}{\hbar} \frac{\partial F(\phi)}{\partial \phi},
\end{equation}
where the free energy is given by \cite{Beenakker:NaMS92}
%OPTION 1:
%\begin{equation}
%    F(\phi)=\int \omega d\omega %\rho_T(\omega, \phi),
%\end{equation}
%and $\rho_T$ is the total DOS.
\begin{equation}
\label{freeF}
    F(\phi) = -2 k_B T \int d\omega \log{\left[2 \cosh{\left(\frac{\omega}{2 k_B T}\right)}\right]} \rho_T(\omega, \phi).
\end{equation}
Here, $\rho_T$ is the total DOS, $k_B$ is the Boltzmann constant and $T$ is the temperature of the system. 
If we assume that the phase variation is mostly localized at the junction, as is typically the case in weak links, this alternative formulation provides an intuitive (although approximate) connection between $J_S$ and the LDOS $\rho$ at the junction. The reason is that the dominant contribution to Eq. \eqref{eq:josF} arises from $\partial \rho /\partial\phi$, see Ref. \cite{Piasotski:PRB24}. Consequently, Eq. \eqref{eq:josF} naturally connects supercurrent $J_S$ and LDOS $\rho$, Eqs. \eqref{eq:LDOS} and \eqref{eq:josK}, which allows to analyze one in terms of the other.

Finally, the critical current is defined as $I_c~=~\text{max}_\phi\left\{J_S(\phi)\right\}$.
%Due to discretization effects inherent in numerical modeling, the junction cannot have strictly zero length. As a result, contributions from above-gap quasi-continuum states must be included when calculating currents \cite{Fatemi:SP25}.
Notice that our model includes the contribution of above-gap states to the supercurrent \cite{Chang:PRB94, Fatemi:SP25}. These cannot be ignored despite being in the short-junction limit because the Andreev approximation (gap much smaller than Fermi energy) does not strictly apply in our models.

\section{Trivial Josephson junctions}
\label{Sec:trivial}

\begin{figure*}
    \centering
    \includegraphics[width=\textwidth]{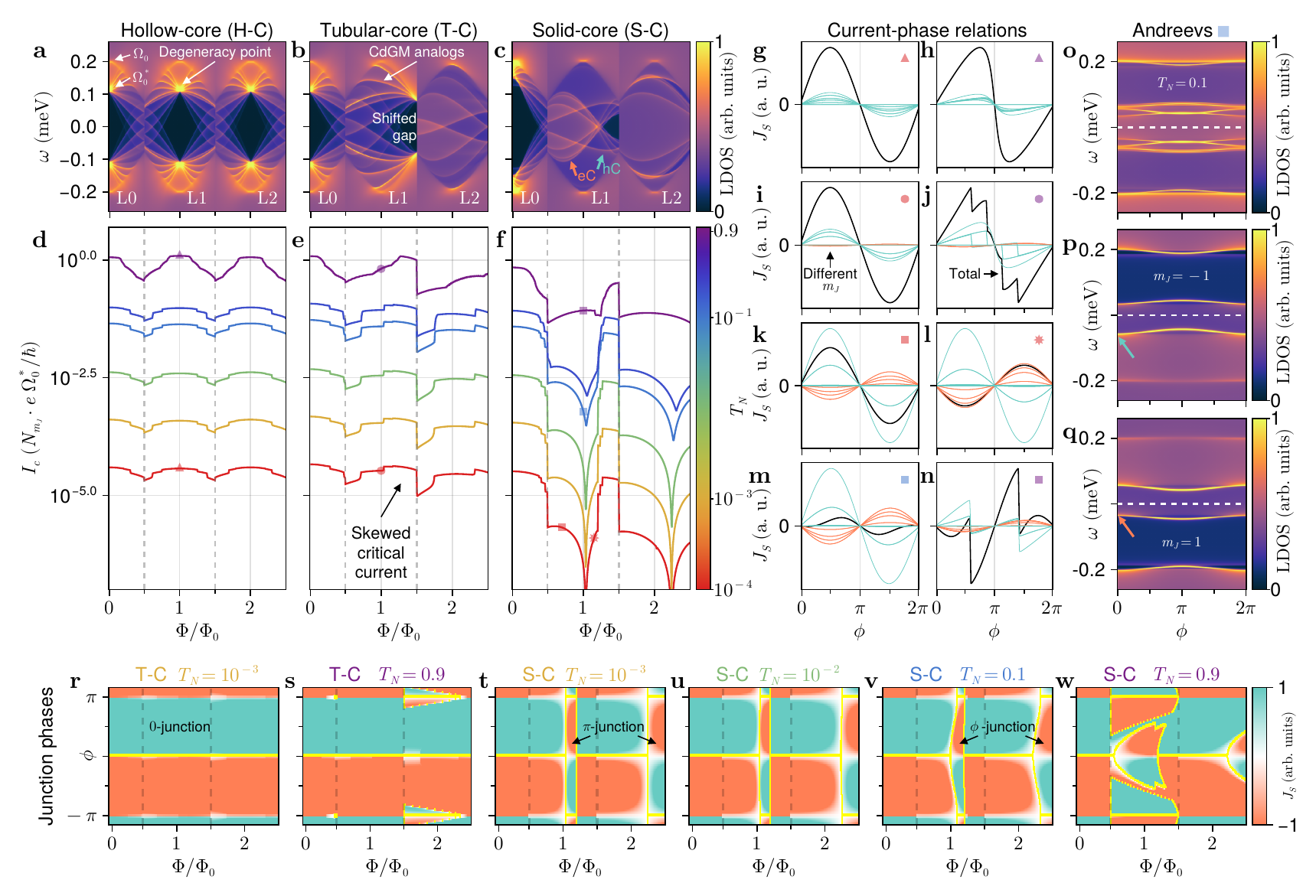}
    \caption{Josephson junctions in the trivial regime. (a-c) Local density of states (LDOS) (in arbitrary units) at the end of a semi-infinite hollow-core, tubular-core and solid-core nanowires, respectively, as a function of energy $\omega$ and applied normalized flux $\Phi/\Phi_0$. The right half of the $n=0$ and the full $n=1,2$ Little-Parks lobes (L0, L1, L2) are displayed. The wire is in the non-destructive LP regime. Degeneracy points, Caroli-de Gennes-Matricon (CdGM) analogs and shifted gaps are highlighted in some panels. $\Omega_0$ and $\Omega^*_0$ are the shell and induced gaps at $\Phi=0$, respectively. (d-f) Critical current $I_c$ (in units of supercurrent unit $e \Omega_0^*/\hbar$ times number of occupied subbands $N_{m_J}$) as a function of $\Phi/\Phi_0$ for different junction normal transmission $T_{\rm{N}}$ and for the different hybrid nanowire models of (a-c). (g-n) Josephson current $J_s$ (in arbitrary units) as a function of junction phase differences $\phi$ for different flux $\Phi$ and transmission $T_{\rm{N}}$ values marked by colored symbols in (d-f). In each panel, the different thin colored curves correspond to the current contributions of different generalized angular momentum $m_J$ filled subbands (turquoise for hole-like CdGM analogs, hCs, and coral for electron-like CdGM analogs, eCs), while the total contribution in shown with a black line. (o) LDOS at the Josephson junction as a function of $\omega$ and $\phi$ for the solid-core model at parameters signaled by the blue square in (f). (p,q) Contribution of the $m_J=-1$ and $m_J=1$ sectors, respectively, to the LDOS in (o). Andreev bound state signals at $\phi\approx 0$ marked by turquoise and coral arrows are related to hCs and eCs in (c). (r-w) 0-, $\pi$- and $\phi$-junction phases versus $\phi$ and $\Phi/\Phi_0$ for different hybrid nanowire models and junction transmissions. Free energy minima are highlighted with thick yellow lines. Parameters: $R=65$nm, $d\rightarrow 0$, $\Omega_0 = 0.23$meV, $\xi_d = 70$nm, $m^*=0.023 m_e$ and $a_0 = 5$nm. For the hollow-core and tubular-core models: $\alpha = 80$meVnm, $g = 0$ and $\Gamma = \Omega_0$ (resulting in $\Omega_0^* \simeq 0.1$meV and $N_{m_J} = 10$). For the hollow-core model: $\mu = 1.5$meV. For the tubular-core model: $W = 20$meV and $\mu = 2$meV. For the solid-core model: $\mu = 9$meV, $\langle \alpha \rangle = 0$meVnm, $U_{\rm{min}} = -30$meV, $U_{\rm{max}} = 0$meV and $\Gamma = 40 \Delta_0$ (resulting in $\Omega_0^* \simeq 0.19$meV and $N_{m_J} = 40$).}
    \label{fig:trivial}
\end{figure*}

We begin our analysis considering short Josephson junctions based on trivial (non-topological) full-shell hybrid nanowires, that is, selecting parameters for which the hybrid nanowires are in the trivial region of the topological phase diagram. For a discussion of the phase diagrams of full-shell nanowires with different core models, see Ref. \cite{Paya:PRB24}. Moreover, we consider that the two superconducting sections at each side of the junction are semi-infinite.

In Figs. \ref{fig:trivial}(a-d) we show the LDOS versus (positive) normalized flux at the end of a single semi-infinite full-shell hybrid nanowire in the trivial regime for different semiconductor-core models, respectively. In the presence of a magnetic flux, the LDOS at low energies features a periodic modulation of the shell gap edge $\Omega(\Phi)$, which gives rise to so-called LP lobes (half of the $n=0$ and the full $n=1,2$ LP lobes are displayed), and a number of subgap features that disperse with flux known as CdGM analogs.  CdGM analogs are shell-induced Van Hove singularities in propagating core subbands, characterized by the generalized angular momentum quantum number $m_J$.  For a fixed flux, they are composed of a bright LDOS peak at a certain subgap energy and a tail that extends towards one (upper or lower) gap edge. The number of CdGM analogs is given by the number of occupied $m_J$ subbands, $N_{m_J}$, which itself depends on $\mu$. We denote $\Omega_0$ ($\Omega^*_0$) the shell (induced) gap at $\Phi=0$. 

In the hollow-core approximation, where all semiconductor charge is assumed to be located as the superconductor-semiconductor interface, the subgap states are very similar for all LP lobes and the CdGM analogs coalesce in the center of the lobes into \textit{degeneracy points} \cite{San-Jose:PRB23}. In this approximation, the induced gap $\Omega^*(\Phi)$ is maximum at the center of the lobes, and is given precisely by the degeneracy-point energy, decreasing towards the edges of the lobes, where CdGM analogs typically cross zero energy. In the tubular-core model, where the charge is distributed across a finite semiconductor tubular section away from the superconductor-semiconductor interface, the degeneracy points of nonzero LP lobes shift towards larger values of magnetic flux (and decrease in energy due to the repulsion with the shell gap edge), the more the larger the lobe number $n$. This produces two important effects: a right skewness of the CdGM analogs and a shifted gap toward the edge of the right lobe (for positive $n$). The degeneracy point and the shifted gap are still visible in the $n=1$ lobe of Fig. \ref{fig:trivial}(b), but have disappeared in the $n=2$ lobe. In the case of a solid-core model, Fig. \ref{fig:trivial}(c), the charge distribution follows the electrostatic dome profile of Fig. \ref{fig:sketch}(b). The LDOS is similar to that of the tubular-core nanowire, especially for steep dome profiles with a depleted core. When this profile is not that steep, the degeneracy points smear out, the amount of right shift is different for each CdGM analog and, in general, a finite CdGM LDOS background covers all nonzero LP lobes \cite{San-Jose:PRB23}.

For the sake of the following discussion, it is worth distinguishing between two types of CdGM analogs. Note that the LDOS is $\pm\omega$ symmetric because of the electron-hole symmetry of the BdG Hamiltonian. The Fermi energy lies at $\omega=0$, meaning that all Bogoliubov states with $\omega\le 0$ are occupied at zero temperature. Focusing thus on the negative energy sector of the $n=1$ lobe of Fig. \ref{fig:trivial}(b), for instance, we see that we have a fan of CdGM analogs dispersing with flux that come directly from the $\omega<0$ degeneracy point that lies at the right lobe edge. These CdGM analogs have gaps extending toward positive energies and Van Hove tails oriented toward negative energies. They originate from hole-like (negative mass) core subbands versus $k_z$. We will call them \textit{hole-Carolis} or hCs.  On the other hand, we have a few CdGM analogs that come from the $\omega>0$ degeneracy point and that have crossed zero energy at some flux into the negative energy sector. They have their gaps extending toward negative energies and their Van Hove tails oriented towards positive energies. They originate from electron-like (positive mass) core subbands versus $k_z$. We will call them \textit{electron-Carolis} or eCs. At low energies, the dispersion of hCs and eCs with flux is opposite. Depending on the core model and the lobe number, hCs or eCs dominate for $\omega<0$, see, for instance, the different lobes of Fig. \ref{fig:sketch}(c). These two types of CdGM analogs are marked with turquoise (hC) and coral (eC) arrows.

In Figs. \ref{fig:trivial}(d-f) we show the associated critical currents $I_c$ to the above three models through a Josephson junction for different junction transparencies $T_N$ \footnote{Note that, strictly speaking, the phenomenology of the critical current can be clearly associated to the LDOS of the decoupled nanowire presented in Figs. \ref{fig:trivial}(a-c) only in the tunneling limit. For high-transparency junctions, deviations exist that could be understood by plotting the junction LDOS at finite $T_N$ and at finite phase difference.}. We normalize $I_c$ to the supercurrent unit $e\Omega_0^*/\hbar$ multiplied by the number of occupied subbands, $N_{m_J}$. In the hollow-core model, Fig. \ref{fig:trivial}(d), $I_c$ is periodically modulated with flux following the LP effect of the shell. It is highest at the center of the lobes, where the induced gap is maximum, and decreases toward the lobe edges in finite-height steps \cite{Giavaras:PRB24} at small transparencies. These steps signal each time an hC CdGM analog crosses over to positive energy and is replaced by an eC crossing into negative energy. When this happens, the hC contribution to the supercurrent is lost and is replaced by a contribution from the eC of opposite sign. The fact that eC contributions to $J_S$ are negative ($\pi$-junction like) will be demonstrated below.

The magnitude of the critical current is proportional to $T_N$, but otherwise its qualitative behavior versus flux is quite independent of junction transparency (except at large transparencies where the steps get rounded; see purple curve). In the tubular-core case, \ref{fig:trivial}(e), the behavior of the critical current is similar, except that now there is a right skewness of $I_c$ versus flux within each $n\neq 0$ LP lobe. Finally, within a solid-core model we find a general behavior similar to that of the tubular core but with a stronger right skewness.

The solid-core model may exhibit an additional and striking phenomenon for certain nanowire parameters, consisting of sharp dips in $I_c$ at particular flux values. In Fig. \ref{fig:trivial}(f) we show one such case. Notice two pronounced dips in $I_c$ around the middle of the $n=1$ and $n=2$ LP lobes, especially at low transparencies. This phenomenon reflects a total or partial suppression of the Josephson current $J(\phi)$ for all $\phi$ simultaneously. It can be understood by inspecting the different contributions to $J(\phi)$ coming from different $m_J$ sectors.

Let us focus on the $n=1$ LP lobe. In Figs. \ref{fig:trivial}(g-n) we show the current-phase relation (CPR) $J_S(\phi)$, in arbitrary units, for flux and transparency values marked by colored symbols in Figs. \ref{fig:trivial}(d-f). In the hollow-core model at low transparencies $T_N\rightarrow 0$, red triangle, we have a typical sine-like tunnel CPR, $J_S(\phi) = I_c\sin(\phi)$; see the black curve in Fig. \ref{fig:trivial}(g). The contributions from different filled $m_J$ subbands, shown with thin curves, are also sine-like, $J_S^{m_J}(\phi)=I_c^{m_J}\sin(\phi)$. In this case, all $I_c^{m_J}$ coefficients are positive. We denote contributions with  $I_c^{m_J}>0$ with turquoise color. They originate from hCs, which favor a conventional 0-junction behavior, since the corresponding contributions to the free energy, $F^{m_J}(\phi)=-E_J^{m_J}\cos(\phi)$, have a positive $E_J^{m_J}=\frac{\hbar}{2e}I_c^{m_J}$ and hence a minimum at $\phi=0$. As the junction approaches perfect transmission, purple triangle, a sawtooth-like CPR emerges; see Fig. \ref{fig:trivial}(h). A similar behavior for low and large $T_N$ is obtained for the tubular-core Josephson junction, Figs. \ref{fig:trivial}(i,j), although now we observe also a small and opposite-sign $I_c^{m_J}<0$ contribution coming populated eCs, shown with thin coral lines. At low transparencies, these eC contributions therefore push the system towards a $\pi$-junction behavior. In the $T_N=0.9$ case (purple circle), the total $J_S$ exhibits also sharp discontinuities against $\phi$ that arise from the abrupt steps in $J_S^{m_J}(\phi)$ whenever a CdGM analog crosses zero energy. Again, a more sophisticated behavior is obtained for the solid-core model. In particular, it exhibits the $I_c$ dips mentioned above, whose origin we can now trace to the competition between eCs and hCs. In Fig. \ref{fig:trivial}(k), we show results for a flux value to the left of the $I_c$ dip in Fig. \ref{fig:trivial}(f), red square. Now, the contribution of the eCs is larger, but the hC contributions still dominate, so that the total $J_S(\phi)$ shows 0-junction behavior. However, to the right of the dip, Fig. \ref{fig:trivial}(f), red star, the eC contributions dominate, and the total CPR switches to a $\pi$-junction behavior \footnote{Note that the flux-driven competition between 0-$\pi$ phases discussed here is unrelated to other previously reported in multicomponent superconductors with doubly-connected cylindrical symmetry \cite{Yerin:PRB22}}; see the black curve in Fig. \ref{fig:trivial}(l). Precisely at the $I_c$ dip, the competing eC and hC contributions cancel. A perfect cancellation for all $\phi$ requires all contributions to be sine-like, i.e. low transparencies. At higher transparencies, $\phi$-junction and more sophisticated phases become possible, as shown in Fig. \ref{fig:trivial}(m,n), corresponding to parameters labeled with blue and purple squares in Fig. \ref{fig:trivial}(f).

\begin{figure*}
    \centering
    \includegraphics[width=\textwidth]{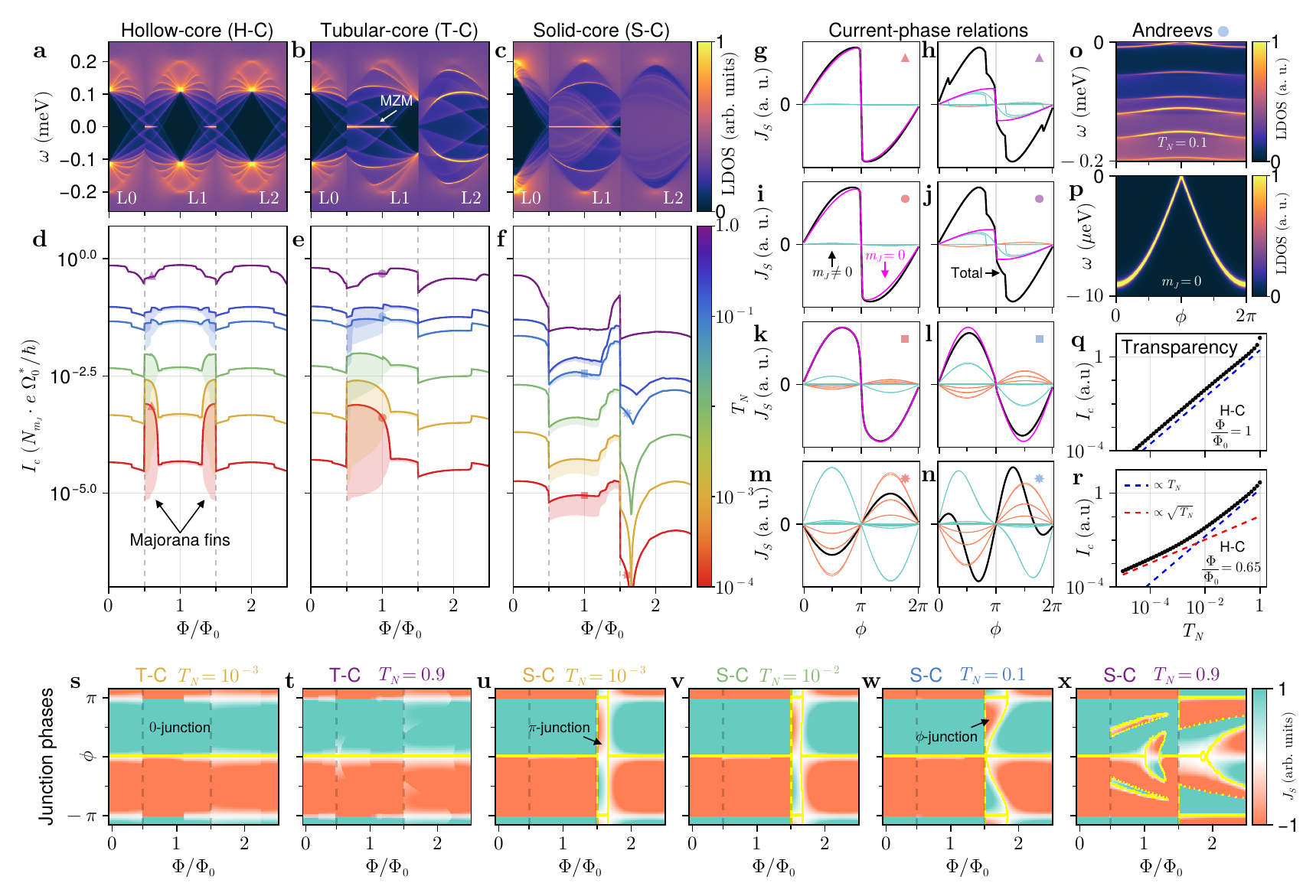}
    \caption{Same as Fig. \ref{fig:trivial} but for Josephson junctions in the topological regime. In (a-c) there appears zero-energy signals in the L1 lobes coming from Majorana zero modes (MZMs). In (d-f), the contribution to the critical current $I_c$ coming from the MZMs is highlighted by colored shadows. This contribution has the shape of ``fins" in the hollow-core and tubular-core cases at low transparencies. In (g-l), the contribution to the Josephson current of the $m_J=0$ sector is shown with a pink curve. (o-r) Critical current $I_c$ versus normal transparency $T_{\rm{N}}$ for different semiconductor-core models at various fluxes. Parameters like in Fig. \ref{fig:trivial} except for $d=5$nm in the tubular- and solid-core models, $\mu = 0.87$meV in the tubular-core model and $\mu = 2$meV $\langle \alpha \rangle = 20$meVnm in the solid-core model ($\Omega_0^*$ and $N_{m_J}$ remain as in Fig. \ref{fig:trivial}).}
    \label{fig:topo}
\end{figure*}

Following the discussion after Eq. \eqref{eq:josF}, the opposite contributions of hCs and eCs to the Josephson effect can be understood by inspecting the LDOS at the junction as a function of $\omega$ and $\phi$. The total junction LDOS for the blue-square case of Fig. \ref{fig:trivial}(f) is shown in Fig. \ref{fig:trivial}(o). In Figs. \ref{fig:trivial}(p,q) we show the contribution of only the $m_J=-1$ and $m_J=1$ sectors, respectively. For $\phi\neq 0$ we observe Andreev states shifting into the CdGM gap. These Andreev states disperse versus $\phi$ with opposite slope for hCs (with $m_J=-1)$ and eCs (with $m_J=1$), since hCs (eCs) have their energy gap oriented towards positive (negative) energies. The contribution to $J_S \propto dF/d\phi$ of hCs and eCs is thus opposite.

Finally, in Figs. \ref{fig:trivial}(r-w) we plot the Josephson-junction current $J_S$ versus flux $\Phi/\Phi_0$ and phase difference $\phi$, where turquoise (coral) color corresponds to a positive (negative) current, and white to a zero current. Free-energy minima (defined by $J_S(\phi) = 0$ with $\partial J_S(\phi)/\partial\phi>0$) are reached at $\phi=\phi_0\in[-\pi,\pi]$ values highlighted with a thick yellow line. These panels show the Josephson phase character versus flux for the different models. The hollow and tubular-core models present a dominant $0$-junction phase. The solid-core model, however, displays $0$- and $\pi$-junction phases to the left and right of the current dips for small transparencies, and $0$- and $\phi$-junction phases at intermediate transparencies. At high transparencies, more sophisticated phase behaviors can develop; see Figs. \ref{fig:trivial}(n,w).

%In summary, the critical current closely follows the LDOS phenomenology of the two semi-infinite full-shell hybrid nanowires that compose the Josephson junction. On the one hand, it is flux-modulated into lobes due to the LP of the shell. On the other hand, within each lobe, $I_c$ depends on the number of filled CdGM analogs and whether they are of the hC or eC character. It is highest at the induced gap maximum (which is core-model dependent), presents steps whenever CdGM analogs in LDOS cross zero energy and, occasionally, presents dips at certain flux values whenever the hC and eC contributions cancel each other. The skewness versus flux of the CdGM analogs in the LDOS translates into a skewness of $I_c$ in non-zero LP lobes, both in tubular and solid-core models. Depending on the prevalence of hC or eC contributions to $J_S$ and the presence of zero-energy crossings, 0, $\pi$ or \editE{ $\phi$-junction} phases can appear in the CPRs.

\section{Topological Josephson junctions}
\label{Sec:topological}

We continue our analysis by considering Josephson junctions based on topological full-shell hybrid nanowires, that is, selecting parameters for which the hybrid nanowires are in the topological region of the topological phase diagram. This means that at each side of the junction there is a Majorana state bound to the end of its corresponding semi-infinite superconducting section. Finite-length hybrid nanowires will be considered in Sec. \ref{Sec:finite}.

A similar analysis to that of Fig. \ref{fig:trivial} is performed for topological junctions in Fig. \ref{fig:topo}. The LDOS versus flux at the end of a single semi-infinite full-shell hybrid nanowire for the three nanowire models, Figs. \ref{fig:topo}(a-c), is very similar to the one in the trivial phase except that now zero-energy peaks (ZEPs) appear for certain flux intervals. Their phenomenology was extensively studied in Ref.~\cite{Paya:PRB24}. It suffices to say here that these ZEPs are signals of MZMs that arise at topological phase transitions of the hybrid wire versus flux. MZMs are possible in the $m_J=0$ sector, and thus they generally appear in odd LP lobes \footnote{Departures from cylindrical symmetry allow for MZMs to emerge also in even lobes \cite{Vaitiekenas:S20,Paya:PRB24}}. Their flux position and extension depend on the wire core model, and hence on the charge distribution across the wire section. In the hollow-core approximation, Fig. \ref{fig:topo}(a), two ZEP intervals appear at the edges of the $n=1$ LP lobe. In the tubular-core model, Fig. \ref{fig:topo}(b), as the degeneracy points shift to larger values of magnetic flux, the right ZEP interval disappears, whereas the left one grows. In the solid core model, Fig. \ref{fig:topo}(b), the ZEP extends throughout the LP lobe.

The critical current associated with the three models is presented in Figs. \ref{fig:topo}(d-f). Its phenomenology is very similar to the trivial case, but now an extra contribution to $I_c$ appears at the flux intervals where MZMs are present, that is added to the one coming from the CdGM analogs. Said contribution appears in the form of what we call \textit{Majorana fins}; see for instance the two thin fin-shaped peaks at the edges of the $n=1$ lobe for the tunnel (red) curve of Fig. \ref{fig:topo}(d), or the single wider fin of Fig. \ref{fig:topo}(e). The Majorana contribution is stronger for smaller transparencies and gradually decreases as the short junction tends to the transparent limit. The magnitude of the MZM contribution to the critical current is highlighted with colored shadows in each $I_c$ curve of Figs. \ref{fig:topo}(d-f). At low transparencies, the current boost is significant. For instance, in the red curve of Fig. \ref{fig:topo}(e), $I_c$ increases from a baseline of $\sim 1$pA in the region without MZMs to $\sim 0.5$nA at $\Phi = 0.5\Phi_0$, i.e., by three orders of magnitude. Notice that in the solid-core model, Fig. \ref{fig:topo}(f), the Majorana contribution is less obvious because the Majorana ZEP covers the whole lobe (so no fin shape is possible) and, additionally, it is comparatively smaller due to the larger number of populated CdGM subbands. Moreover, the $I_c$ dips in the $n=1$ LP lobe of Fig. \ref{fig:trivial}(f) are now overshadowed by the Majorana contribution, which cannot cancel out.

The above phenomenology can again be understood by inspecting the CPRs of the three models at different values of $\Phi$ and $T_N$. This analysis is presented in Figs. \ref{fig:topo}(g-n). We now highlight the $m_J=0$ contribution to $J_S(\phi)$ with a thin pink curve. The Majorana contribution in said sector is always of the 0-junction type and much larger than that of the different CdGM analogs in the tunneling limit. Ultimately, this comes from the way that the two Majorana bound states at the ends of the superconducting sections hybridize through the junction as a function of $\phi$. The $m_J=0$ LDOS at low energies versus $\phi$ is shown in Fig. \ref{fig:topo}(p). The Andreev state from the hybridization of the two MZMs at the junction exhibits a protected zero-energy crossing at $\phi=\pi$ (to be compared with the behavior of all other CdGM analog subbands, whose energy is in general finite). This is the origin of the $4\pi$-periodic Josephson effect of closed topological Josephson junctions when fermion parity is fixed. In our case, we are considering open Josephson junctions in equilibrium, allowing to change parity at $\phi=\pi$. As a result, the pink curves in Figs. \ref{fig:topo}(g-l) are $2\pi$-periodic. Still, the Majorana contribution remains unusual, since it is proportional to $\sqrt{T_N}$ \cite{Cayao:PRB17,San-Jose:PRL14}, as corresponds to half-quasiparticles, instead of proportional to $T_N$, as is the case for all other hC and eC quasiparticles. This behavior is demonstrated for the hollow core model in Figs. \ref{fig:topo}(q,r). In the center of the $n=1$ LP lobe, where there are no Majoranas, $I_c$ grows linearly with transmission. However, at $\Phi/\Phi_0=0.65$, where the Majorana contribution is present, the current scales at low transparencies as $\sqrt{T_N}$. This same scaling has been verified for the tubular- and solid-core models (not shown).  Finally, the junction phases are shown in Figs. \ref{fig:topo}(s-x).

%In summary, the presence of MZMs at the weak-link junction produces an excess critical-current contribution to be added to that coming from trivial CdGM analogs. This Majorana contribution is more visible at low junction transparencies $T_N$, since it scales as $\sqrt{T_N}$, compared to the linear-in-$T_N$ one of trivial states. At low transparencies, therefore, the presence of MZMs could be detected through $I_c$ measurements, particularly in core models where they produce fin-like peaks in $I_c$ along finite flux intervals.

\section{Finite-length hybrid nanowires}
\label{Sec:finite}

\begin{figure*}
    \centering
    \includegraphics[width=\textwidth]{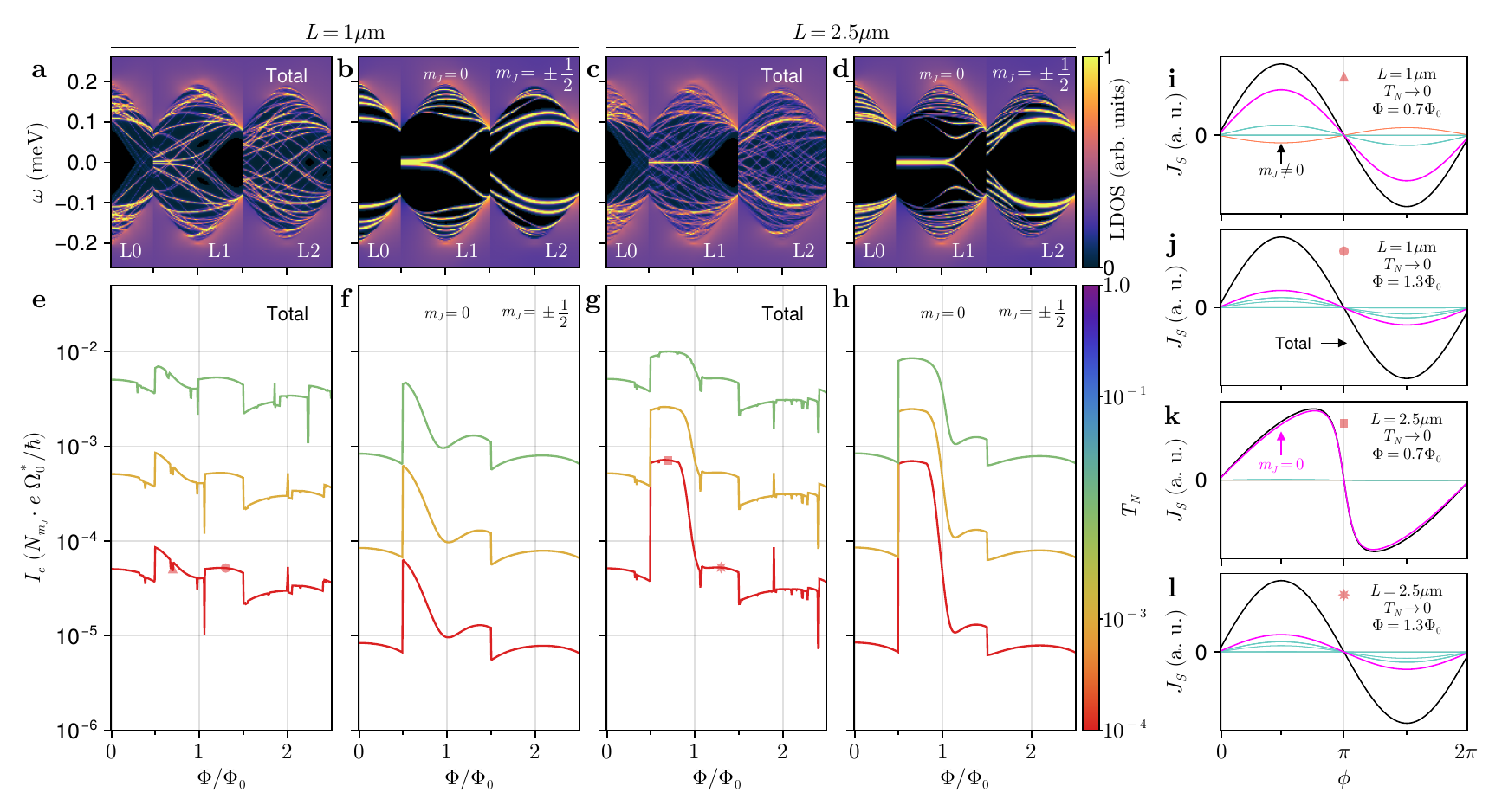}
    \caption{Topological Josephson junctions with finite-length full-shell hybrid nanowires. (a) LDOS (in arbitrary units) at one end of a $L=1\mu$m tubular-core nanowire as a function of energy $\omega$ and applied normalized flux $\Phi/\Phi_0$. (b) Contribution of the lowest $|m_J|$ sector in each lobe to the total LDOS displayed in (a). (c,d) Same as (a,b) but for a $L=2.5\mu$m tubular-core nanowire. (e-h) Critical current $I_c$ as a function of $\Phi/\Phi_0$ for different junction normal transmission $T_{\rm{N}}$ and for the different hybrid nanowire cases of (a-c), respectively. (i-l) Current-phase relations in the tunneling regime for parameters marked by red symbols in (e) and (g). Parameters as for the tubular-core model of Fig. \ref{fig:topo}.}
    \label{fig:flength}
\end{figure*}

We finish by considering topological Josephson junctions between finite-length full-shell hybrid nanowires. Now, Majorana bound states appear at both ends of each superconducting section, so that there are four Majoranas in total; see Fig. \ref{fig:sketch}(c).
%Still, the ones that overlap the more through the junction are the inner Majoranas.
When the length of each hybrid nanowire, $L$, is much larger than the Majorana localization length, $\xi_M$, the left and right Majoranas are effectively decoupled and remain at zero energy. However, when $L\lesssim 2\xi_M$, the left and right Majoranas overlap. We analyzed in a previous work~\cite{Paya:PRB24a} the phenomenology of a single finite-length full-shell hybrid nanowires and we found that, for realistic parameters, there are in general no Majorana oscillations versus flux in the LDOS  of Al/InAs-based nanowires. The reason is simply that the lobe flux interval is too small for complete oscillations to develop. Thus, in general, MZMs in finite-length $L\lesssim 2\xi_M$ nanowires split in energy, turning into non-oscillating trivial quasiparticle states.

Focusing on the tubular-core model for concreteness, in Fig. \ref{fig:flength}(a) we show the LDOS versus flux at the end of a full-shell hybrid nanowire of length $L=1\mu$m. The Majorana localization length is generally flux dependent, but in this case, for most of the $n=1$ LP lobe, $300\text{nm}\lesssim\xi_M\lesssim 1000\text{nm}$. Thus, $L\sim 2\xi_M$.  Two important changes occur with respect to the semi-infinite case of \ref{fig:trivial}(b). On the one hand, the different CdGM analogs split into several longitudinally confined levels, densely covering the nonzero LP lobes. On the other hand, the MZM splits in energy for most of the previous ZEP interval. This can be clearly seen in Fig. \ref{fig:flength}(b), where the lowest $m_J$-sector contribution to the total LDOS is shown ($m_J=\pm 1/2$ [$m_J=0$] for even [odd] lobes).

The total critical currents $I_c$ for low transparencies are plotted in Fig. \ref{fig:flength}(e). The different curves are similar in shape and magnitude to the results obtained for the trivial Josephson junction, Fig. \ref{fig:trivial}(e), but now they exhibit noise due to problematic convergence of numerical integrals. Figure \ref{fig:flength}(f) reveals that even if there are still traces of the Majorana fin in the $m_J=0$ sector of the $n=1$ LP lobe, this contribution does not stand out in the total $I_c$.

A longer full-shell hybrid nanowire is considered in Fig. \ref{fig:flength}(c). A finer mesh of CdGM levels is present, as corresponds to a longer longitudinal confinement of the original infinite-length CdGM analogs. Now, the Majorana ZEP remains close to zero energy along the whole original interval; Fig. \ref{fig:flength}(d). Consequently, the presence of MZMs at the junction is again detectable in $I_c$ at small transparencies; Figs. \ref{fig:flength}(g,h). As in Fig. \ref{fig:topo}(e), on the left side of the $n=1$ LP lobe there is a clearly visible fin-shaped extra contribution over the CdGM background. Indeed, analyzing the CPR for the parameters marked by a red square in Fig. \ref{fig:flength}(g), we find $J_S(\phi)$ curves with a strong sawtooth component, which implies that essentially all the critical current at that flux comes from the Majoranas in the $m_J=0$ sector; see Fig. \ref{fig:flength}(k).

%\editE{A similar study on trivial Josephson junctions (not shown) reveals that $I_c$ for finite-length superconducting sections is essentially the same (both in magnitude and shape) as the semi-infinite length version, Fig. \ref{fig:trivial}(d-f).}

\section{Summary and conclusions}
\label{Sec:conclusions}

In this work, we have analyzed the rich phenomenology of short Josephson junctions based on full-shell hybrid nanowires, considering different models for charge distribution inside the semiconductor core and different junction transmissions.

In trivial Josephson junctions, we have seen that the critical current $I_c$ follows the LDOS phenomenology of the isolated superconducting sections that make up the junction. On the one hand, $I_c$ is flux-modulated into lobes due to the LP of the shell. On the other hand, within each lobe, $I_c$ depends on the number of filled CdGM analogs and whether they are of the hC or eC character. $I_c$ is highest at the induced gap maximum (which is core-model dependent), presents steps whenever CdGM analogs in LDOS cross zero energy, and, occasionally, presents dips at certain flux values whenever the hC and eC contributions cancel each other. The skewness versus flux of the CdGM analogs in the LDOS translates into a skewness of $I_c$ in nonzero LP lobes \footnote{Hints of critical current skewness (or more specifically, skewness of switching currents) have been measured in Refs. \cite{Razmadze:PRB24a, Ibabe:NL24}.}, both in tubular- and solid-core models. Depending on the prevalence of hC or eC contributions to $J_S$, $0$-, $\pi$- or $\phi$-junction phases can appear in the CPRs, with the latter requiring medium-to-high transparencies.

In topological Josephson junctions with long superconducting sections, the presence of MZMs at the weak link produces an excess critical-current contribution that adds to that coming from trivial CdGM analogs. This Majorana contribution is visible at low junction transparencies since it scales as $\sqrt{T_N}$, compared to the linear-in-$T_N$ one of trivial states. At low transparencies, therefore, MZMs could be detected through $I_c$ measurements versus flux in long wires with depleted cores, where they manifest as fin-like peaks. A way to test whether a peak in $I_c$ in a flux interval of the $n=1$ LP lobe has a Majorana origin would be to change the transparency of the weak link, e.g. through a plunger gate at the junction. The critical current should crossover from a $\sqrt{T_N}$ to a $T_N$ behavior as the junction transparency increases. Alternatively, a dominant MZM contribution could be demonstrated by a sawtooth-like CPR even at low transparencies, with finite jumps at phase $\pi$. Both the fin-like $I_c$ peaks and the CPR jumps are expected to survive up to temperatures comparable to the Majorana hybridization across the junction, i.e. for $k_BT\lesssim \Omega_0^* \sqrt{T_N}$.
%If the peak essentially remains, like the one at the right edge of the first lobe in Fig. \ref{fig:topo}(f), it is probably originating from a skewed gap. However, the Majorana peak should disappear

When the length of the superconducting sections is of the order of or smaller than the Majorana localization length $\xi_M$, $I_c$ behaves essentially as in trivial Josephson junctions. Thus, it is necessary to have $L\gg\xi_M$ to clearly observe Majorana signatures in $I_c$. Moreover, to observe Majorana physics, it is also necessary to have sufficiently clean full-shell hybrid nanowires and junctions, as we consider here. The detrimental effects of disorder of various types on Majorana nanowires have been extensively analyzed before \cite{Kells:PRB12,Prada:PRB12,Rainis:PRB13,Roy:PRB13,Liu:PRB17a,Reeg:PRB17,Penaranda:PRB18,Vuik:SP19,Rossi:PRB20,Prada:NRP20,Pan:PRR20,Woods:PRA21,DasSarma:PRB21,DasSarma:PRB23,Hess:PRL23,Pan:PRB24,Taylor:PRL24, Paudel:PRA25}, and we expect similar conclusions here. Concerning Josephson junctions in the trivial regime, it is interesting to note that the critical current remains essentially unaffected for finite-length full-shell nanowire sections as compared to semi-infinite ones. This points to a notable resilience of the contribution of CdGM analogs to $I_c$ with respect to strong disorder along the wires (the type of disorder that effectively splits the superconducting sections into finite-length pieces).

%Apart from potential Majorana physics, we note that the phenomenology of these Josephson junctions in the trivial regime is remarkable, as it is dominated by LP and CdGM analog physics. 
%Since the LP effect is a property of the (diffusive) superconductor shell, and the CdGM analogs are delocalized quasi-particle states (as opposed to zero-energy nonlocal Majorana bound states), we expect their signatures in $I_c$ to be more resilient against disorder \cite{Deng:PRL25}. \editE{For instance, CdGM analogs have been observed in Ref. \cite{Deng:PRL25},  Ultimately, if the disorder is so strong that it effectively breaks the nanowires into several finite-length sections, we expect similar results for $I_c$ to the ones discussed in Sec. \ref{Sec:finite}.

We have focused for concreteness on full-shell hybrid nanowires in the non-destructive LP regime. However, our conclusions are trivially generalized to the destructive regime case. In this case, the critical current vanishes between lobes as the shell gap closes. In addition, for simplicity, we have considered short junctions, where the phenomenology is already quite complex. Going beyond the short-junction limit, even richer supercurrent behaviors could be explored, including for example all the physics that arises in junctions containing quantum dots \cite{Valentini:S21,Razmadze:PRL20,Escribano:PRB22,Svetogorov:PRB23}. Lastly, in this work we have concentrated on Josephson junctions with equal superconducting sections. When the characteristics of the two sections are sufficiently different, as for example for full-shell hybrid nanowires with different radii or different superconducting coherence lengths, new effects appear, including a fluxoid valve effect controlled by the applied magnetic field \cite{Paya:25d}.

Finally, we believe that our calculations provide a versatile starting point for theoretical modeling of novel concepts and devices based on full-shell Josephson junctions. In this context, systems of great current interest include superconducting qubits based on full-shell nanowires \cite{Sabonis:PRL20,Kringhoj:PRL21,Erlandsson:PRB23} where, arguably, the full potential offered by their flux-tunability, beyond demonstrating flux-dependent qubit frequencies, has not been fully exploited. This involves exploring the $0$, $\pi$ and $\phi$ regimes discussed here, as well as the possibility of tailoring the high-harmonics content of the Josephson potential in order to reach protected $\cos 2\phi$ qubit regimes \cite{Giavaras:25}. From a broader perspective, it is important to note that our Green's function approach in Eq. \eqref{eq:josK}, or alternatively a direct evaluation of the Josephson potential from the free energy in Eq. \eqref{freeF}, would allow us to go much deeper into novel areas of research where theoretical studies beyond simple short junction Andreev models \cite{Beenakker:TPMS92} are almost non-existent. This includes non-Hermitian Josephson junctions \cite{Cayao:PRB24a,Shen:PRL24,Beenakker:APL24,Pino:PRB25} and multiterminal Josephson junction geometries, that allow the study of high-dimensional synthetic bands exhibiting a wide range of new effects including topological Andreev bands and Weyl nodes \cite{Riwar:NC16,Coraiola:NC23,Matute-Canadas:PQ24,Antonelli:25}.

All the numerical codes used in this paper were based on the Quantica.jl package \cite{San-Jose:25}. The specific code to build the nanowire Hamiltonian and to perform and plot the calculations is available at Refs. \cite{Paya:25a} and \cite{Paya:25e}, respectively. Visualizations were made with the Makie.jl package \cite{Danisch:JOSS21}. 
\acknowledgments{
This research was supported by Grants PID2021-122769NB-I00, PID2021-125343NB-I00 and PRE2022-101362 funded by MICIU/AEI/10.13039/501100011033, ``ERDF A way of making Europe'' and ``ESF+''. Part of this research project was made possible through the access granted by the Galician Supercomputing Center (CESGA) to its supercomputing infrastructure.
}

\appendix

%\editE{\section{Nanowire Hamiltonian}}
\section{Nanowire Hamiltonian}
\label{ap:hamiltonian}

This appendix summarizes the key features of the full-shell nanowire models relevant to this work. These models are thoroughly described and discussed in Refs. \cite{San-Jose:PRB23, Paya:PRB24}. 

A full-shell nanowire consists of a semiconductor core proximitized on all facets by a diffusive superconducting shell. By integrating out the shell degrees of freedom, we obtain a self-energy $\Sigma_{\rm shell}$ acting on the core surface. The Green's function is then given by $G(\omega) = \left[\omega - H_{\rm core} - \Sigma_{\rm shell}(\omega) \right]^{-1}$, where $ H_{\rm core}$ is the Hamiltonian of the core. For analytical convenience, we express the model in terms of an effective Bogoliubov-de Gennes (BdG) Hamiltonian $H \equiv \omega - G^{-1}(\omega) = H_{\rm core} + \Sigma_{\rm shell}$, expressed in the Nambu basis $\Psi = \left(\psi_\uparrow, \psi_\downarrow, \psi^\dagger_\downarrow, -\psi^\dagger_\uparrow \right)$. 

Assuming a cylindrical symmetry for the full-shell nanowire [with coordinates $(r,\varphi,z)$], it can be shown that $\left[H, J_z\right] = 0$, where the generalized angular momentum $J_z = -i \partial_\varphi + \frac{1}{2}\sigma_z + \frac{1}{2}n \tau_z$ is the sum of the orbital angular momentum, spin momentum and ``fluxoid'' momentum. Here, $\sigma_i$ ($\tau_i$) are Pauli matrices for the spin (particle-hole) space, and $n$ is the fluxoid number. We can then block-diagonalize $H$ on subspaces with good quantum numbers $m_J$, the eigenvalues of $J_z$. These quantum numbers take values given in Eq. \eqref{mJ} \cite{Vaitiekenas:S20}.
On $m_J$ subspaces, $H$ takes a $\varphi-$independent form,
\begin{equation}
    \begin{split}
        \tilde{H} =& \left[\frac{p_z^2 + p_r^2}{2 m^*} + U(r) - \mu\right]\sigma_0 \tau_z + V_Z \sigma_z \tau_0 \\
        &+ \frac{1}{2 m^* r^2}\left(m_J - \frac{1}{2}\sigma_z - \frac{1}{2}n\tau_z + \frac{1}{2}\frac{\Phi}{\Phi_0} \frac{r^2}{R_{\rm LP}^2}\tau_z\right)^2\sigma_0\tau_z \\
        &- \frac{\alpha(r)}{r}\left(m_J - \frac{1}{2}\sigma_z - \frac{1}{2}n\tau_z + \frac{1}{2}\frac{\Phi}{\Phi_0} \frac{r^2}{R_{\rm LP}^2}\tau_z\right)\sigma_z\tau_z \\
        & + \alpha(r)k_z\sigma_y\tau_z + \Sigma_{\rm shell}(\omega),
    \end{split}
\end{equation}
where $R_{\rm LP} = R + d/2$, $p_z = -i \partial_z$, $p_r = -i \partial_r$, $p_r^2 = -\frac{1}{r}\partial_r \left(r \partial_r\right)$, $m^*$ is the semiconductor effective mass, $\mu$ the chemical potential (with $\hbar = 1$) and $\Phi=\pi R_{\rm{LP}}^2B$ is the flux, where $B$ is the axial magnetic field. The Zeeman field is included as $V_Z = \frac{1}{2}g \mu_B B$, where $g$ is the Landé factor and $\mu_B$ the Bohr magneton. $U(r)$ denotes the electrostatic potential inside the core. Although the precise form of $U(r)$ depends on the microscopic details of the interface, it is known that an epitaxial core/shell Ohmic contact leads to a dome-shaped profile \cite{Mikkelsen:PRX18, Antipov:PRX18}, which we model as
\begin{equation}
    U(r) = U_{\rm min} + \left(U_{\rm max} - U_{\rm min}\right) \left(\frac{r}{R}\right)^2.
\end{equation}
We assume a Rashba type SOC arising from the inversion symmetry breaking at the core/shell interface, which is radial and points outwards, $\alpha(r) = -\alpha_0 \partial_r U(r)$, with $\alpha_0$ a model parameter.

The self-energy of the superconducting shell is written in terms of a normal-state decay rate $\Gamma$ into the shell. Following Ref. \cite{Skalski:PR64},
\begin{equation}
    \Sigma_{\rm shell}(\omega) = \Gamma \sigma_0 \frac{\tau_x - u(\omega)\tau_0}{\sqrt{1 - u(\omega)^2}},
\end{equation}
where $u(\omega)$ satisfies
\begin{equation}
    u(\omega) = \frac{\omega}{\Delta (\Lambda)} + \frac{\Lambda}{\Delta (\Lambda)} \frac{u(\omega)}{\sqrt{1-u(\omega)^2}},
\end{equation}
and  $\Lambda$ is a depairing parameter. The pairing  $\Delta$ obeys
\begin{equation}
    \begin{split}
        \ln\frac{\Delta(\Lambda)}{\Delta(0)} =& -P\left(\frac{\Lambda}{\Delta(\Lambda)}\right),\nonumber\\
P(\lambda\leq 1) =&\frac{\pi}{4}\lambda,\nonumber\\
P(\lambda\geq 1) =& \ln\left(\lambda+\sqrt{\lambda^2-1}\right)+\frac{\lambda}{2}\arctan\frac{1}{\sqrt{\lambda^2-1}}\nonumber\\
&-\frac{\sqrt{\lambda^2-1}}{2\lambda}.
    \end{split}
\end{equation}
We select the solution for $u(\omega)$ that ensures the correct continuity and asymptotic behavior of the retarded Green's function. The superconducting gap is given by \cite{Skalski:PR64}
\begin{equation}
    \Omega(\Lambda) = \left(\Delta(\Lambda)^{2/3} - \Lambda^{2/3}\right)^{3/2}.
\end{equation}
The relation between the depairing parameter and the magnetic flux is \cite{Schwiete:PRB10}
\begin{equation}
   \begin{split}
       \Lambda(\Phi) &= \frac{k_{B} T_{\rm c}\,\xi_{d}^2}{\pi R_{\rm{LP}}^2}\left[4\left(n-\frac{\Phi}{\Phi_0}\right)^2 + \frac{d^2}{R_{\rm{LP}}^2}\left(\frac{\Phi^2}{\Phi_0^2} + \frac{n^2}{3}\right)\right],\\
    n(\Phi) &= \lfloor \Phi/\Phi_0\rceil = 0, \pm 1,\pm 2, \dots,
   \end{split}
   \label{depairing}
\end{equation}
where $\xi_d$ is the diffusive superconducting coherence length, $T_c$ is the zero-flux critical temperature and $k_B$ the Boltzmann constant.

In the tubular-core model, the wave function is assumed to be radially concentrated within a distance $W$ from the superconductor-semiconductor interface. In this case, we redefine
\begin{equation}
    \begin{split}
        \langle U(r) \rangle - \mu &\rightarrow \mu, \\
        \langle \alpha(r) \rangle &\rightarrow \alpha,
    \end{split}
\end{equation}
i.e., we take constant values for these parameters along $r\in[0,W]$. (Note that in the solid-core model, the full radial dependence of the Hamiltonian is taken into account.) It has been shown that, for $W\lesssim R/2$ and low doping, evaluating the tubular-core Hamiltonian solely at the average radius $R_\text{av}=R-W/2$ preserves accuracy while drastically improving computational efficiency \cite{San-Jose:PRB23,Paya:PRB24}. This is formalized as
\begin{equation}
    \Psi_W(r) = \delta(r - R_\text{av})\Psi,
\end{equation}
reducing the Hamiltonian to $\tilde{H}^{TC} = \mel{\Psi_W}{\tilde{H}}{\Psi_W}$; see Eqs (A15)-(A19) in Ref. \cite{Paya:PRB24a}. The hollow-core model corresponds to $W \to 0$, confining the wavefunction strictly to $r = R$. 

The hollow-core Hamiltonian \cite{Vaitiekenas:S20} and the tubular-core Hamiltonian $\tilde{H}^{TC}$ \cite{Paya:PRB24,Paya:PRB24a} can be mapped to a one-dimensional Majorana nanowire \cite{Lutchyn:PRL10, Oreg:PRL10} for $m_J = 0$. Unlike in the latter, in the full-shell geometry the topological phase transition is driven by an \textit{effective} Zeeman field 
\begin{equation}
V_{\rm Z}^{\Phi} = \frac{1}{2}\left(n(\Phi)- \frac{R_{\rm av}^2}{R_{\rm LP}^2}\frac{\Phi}{\Phi_0}\right) \left( \frac{1}{2 m R_{\rm av}^2} + \frac{\alpha}{R_{\rm av}} \right),
\end{equation}
that depends on the orbital effect of the magnetic flux $\Phi$ and is present even in the absence of semiconductor $g$ factor. The orbital effect allows for the topological transition to take place at far lower magnetic fields than those required in Zeeman-driven, partial-shell Majorana nanowires. In fact, as the magnetic fields considered throughout this work are small, we simplify the calculation by setting the $g$ factor to $0$. Thanks to this mapping, it is possible to understand that when the wire is threaded by an odd number of fluxoids $n$, the $m_J=0$ subband sector may undergo a topological phase transition and develop MZMs depending on the paremeters of the wire (such as SOC $\alpha$ or chemical potential $\mu$). The topological phase diagrams for these Hamiltonians were computed in Ref. \cite{Paya:PRB24}.

\bibliography{Josephson, biblio}

\end{document}